\documentclass[reprint,aps,showpacs,floatfix]{revtex4-1}
\usepackage{graphics,graphicx,epsfig}

\begin{document}

\title{Random sequential adsorption of polydisperse mixtures on lattices}
\author{R.C.Hart}
\email{hart@if.uff.br}

\author {F. D. A. Aar\~ao Reis}
\email{ reis@if.uff.br}
\affiliation{Instituto de F\'\i sica, Universidade Federal Fluminense,
Avenida Litor\^anea s/n, 24210-340 Niter\'oi RJ, Brazil}

\begin{abstract}

Random sequential adsorption of linear and square particles with excluded volume interaction
is studied numerically on planar lattices considering Gaussian distributions of lateral sizes
of the incident particles, with several values of the average $\mu$ and of the width-to-average
ratio $w$.
When the coverage $\theta$ is plotted as function of the logarithm of time $t$, the maximum
slope is attained at a time $t_M$ of the same order of the time $\tau$ of incidence of one
monolayer, which is related to the molecular flux and/or sticking coefficients.
For various $\mu$ and $w$, we obtain $1.5\tau< t_M < 5\tau$ for linear particles and
$0.3\tau < t_M < \tau$ for square particles.
At $t_M$, the coverages with linear and square particles are near $0.3$ and $0.2$, respectively.
Extrapolations show that coverages may vary with $\mu$ up to $20\%$ and $2\%$ for linear and
square particles, respectively, for $\mu\geq 64$, fixed time, and constant $w$.
All $\theta\times \log{t}$ plots have approximately the same shape, but other quantities
measured at times of order $t_M$ help to distinguish narrow and broad incident distributions.
The adsorbed particle size distributions are close to the incident ones up to long times for
small $w$, but appreciably change in time for larger $w$, acquiring a monotonically
decreasing shape for $w=1/2$ at times of order $100\tau$.
At $t_M$, incident and adsorbed distributions are approximately the same
for $w\leq 1/8$ and show significant differences for $w\geq 1/2$; this result may be
used as a consistency test in applications of the model.
The pair correlation function $g\left( r,t\right)$ for $w\leq 1/8$ has a well defined
oscillatory structure at $10t_M$, with a minimum at $r\approx \mu$ and maximum at
$r\approx 1.5\mu$, but this structure is not observed for $w\geq 1/4$.

\end{abstract}


\maketitle

\section{Introduction}
\label{intro}

Models of random sequential adsorption (RSA) describe real processes in which atoms or
molecules sequentially adsorb on a substrate to form a monolayer or a multilayer
\cite{privman,adamczykbook}.
A broad range of applications include adsorption of colloidal particles
\cite{adamczykbook,yuan} and of proteins \cite{rabe,adamczykCurOp} on various substrates,
growth of atomic islands of metals or semiconductors \cite{etb}, functionalization
of semiconductor surfaces with molecular monolayers \cite{hamers2008}, etc.
The RSA models may consider one or more species of adsorbing particles, different particle
shapes, discrete or continuous size distributions, and include other surface processes such
as diffusion and desorption \cite{privman,adamczykbook,evans1993,evans2000,schaaf2000}.

Polydispersity of incident particle size is observed in a large number of processes,
specially in (but not restricted to) colloidal particle and macromolecule deposition.
The simplest RSA models that account for polydispersity effects are those with
binary particle size distribution \cite{doty,loncarevic2007,subashiev,talbot,dias},
which were already used to explain real system properties \cite{qiu,rosenberg}.
Uniform distributions of particle size were also considered in continuum
\cite{danwanichakul,tarjous} and lattice \cite{budinski2008,loncarevic2010,budinski2011}
models, showing interesting features such as the effects of particle shape on the
maximal surface coverage and on the form of concentration decay.
For some applications, models of polydisperse mixtures with power-law size
distributions were also proposed \cite{brilliantov,vieira}.
Surface diffusion and desorption were neglected in most cases. 

RSA of mixtures of spheres and disks with Gaussian size distributions were also studied
in Refs. \protect\cite{meakin,adamczyk1997,marques}.
Some important conclusions from those works were the increase of the
adsorption rate and the shift of the adsorbed particle size distribution peak
as the width of the incident distribution increased.
These results helped to interpret data for polystyrene nanoparticle adsorption
on charged layers over silicon substrates \cite{hanarp}.
Recent works also model adsorption of silver and hematite nanoparticles by RSA of
polydisperse mixtures \cite{ocwieja2012,ocwieja2013,ocwieja2015}. 

The aim of the present work is to study related RSA
problems on lattices, in which submonolayers are formed by adsorption of particles
with sizes following discretized Gaussian distributions.
Two limiting cases of particle shape are considered separately, namely linear segments and
filled squares.
Broad ranges of the average lateral size and of the width-to-average ratio are considered
to model the incident flux.
The evolution of the coverage shows that plots of this quantity as function
of $\log{t}$ have maximal slopes in narrow time ranges which are related to the
incident flux, with weaker effects of particle shape, average size, and distribution width.
This feature may be used, for instance, to estimate orders of magnitude of molecular flux
or sticking coefficients.
The adsorbed particle size distributions are also analyzed and the ranges of time and
distribution widths in which they are similar to the incident ones are presented.
For large widths and times not very long, the adsorbed distributions are monotonically
decreasing.
The pair correlation functions of the adsorbates are also analyzed and may be used
to distinguish cases of narrow and broad incident size distributions.

% small change R1Q1
The rest of this work is organized as follows. In Sec. \ref{models}, we present the RSA models,
information on the simulations, and the quantities to be analyzed.
In Sec. \ref{linear}, we present results for RSA of polydisperse mixtures of linear particles.
In Sec. \ref{square}, we present results for RSA of polydisperse mixtures of squares.
Sec. \ref{conclusion} summarizes our results and present our conclusions.

\section{Model, simulations, and basic quantities}
\label{models}

\subsection{The RSA models}

\begin{figure}[!h]
\centering
\includegraphics[width=8cm]{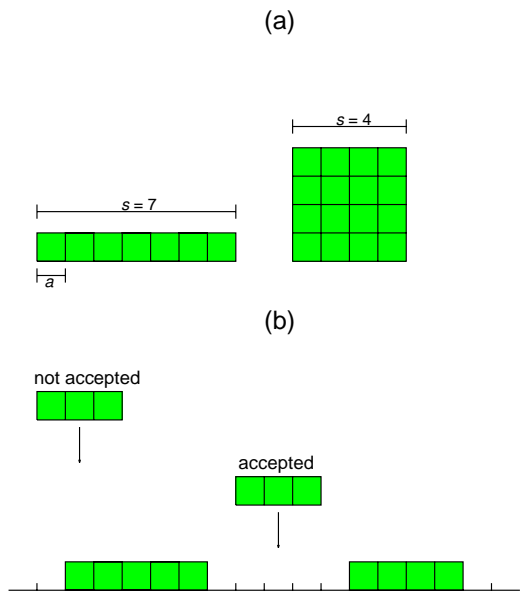}
\caption{(Color online)
(a) Example of linear (left) and square (right) particles with indication of their
dimensionless sizes $s$ and of the size $a$ of the lattice site.
(b) Incidence of linear particles in a partially filled lattice showing cases of
accepted and rejected adsorption trials.
}
\label{model}
\end{figure}

The substrate in which particles adsorb is a square lattice in the $xy$-plane.
The edge of a lattice site is $a$, with corresponding site area $a^2$.
Adsorption of two types of particles are separately considered, namely linear particles and
filled squares, which are illustrated in Fig. \ref{model}a.

The particles sequentially incide at randomly chosen surface positions, moving in the
$z$ direction.
A particle attaches to the surface only if all target sites are empty, i. e. if all positions
$\left( x,y\right)$ occupied by the incident particle correspond to empty sites at the
substrate.
Otherwise, the adsorption attempt is rejected.
Accepted and rejected adsorption attempts on a line are illustrated in Fig. \ref{model}b.

The length of a linear particle is $sa$, where $s$ is the number of sites that it occupies.
The size $s$ is chosen from a Gaussian distribution of average $\mu$ and width $\sigma$:
\begin{equation}
Q\left( s\right) = \frac{1}{\sqrt{2\pi \sigma^2}}
\exp{\left[ -\frac{{\left( s-\mu\right)}^2}{2\sigma^2}\right]} .
\label{distrincident}
\end{equation}
Note that $s$, $\mu$, and $\sigma$ are dimensionless quantities; alternatively, they are
lengths given in units of the lattice constant $a$.
Only integer values of $s$ between $s=1$ and $s=2\mu$ are chosen in our simulations, thus
the normalization factor is slightly different from that in Eq. \ref{distrincident}.
The width-to-average ratio
\begin{equation}
w\equiv \sigma /\mu
\label{defr}
\end{equation}
is the main quantity to characterize the shape of $Q\left( s\right)$ and consequently
the degree of polydispersity.
Hereafter, $Q\left( s\right)$ is simply called incident distribution.

In the deposition of a square particle, a discretization of $Q\left( s\right)$ is also used
to choose the lateral size $s$ of the incident square.
This particle has $s^2$ sites and area $s^2a^2$.

The basic time unit $\tau$ is the time necessary to fill the whole substrate if all
adsorption attempts are accepted.
The complete filling of the substrate would be achieved, for instance, if the incidence of
particles was ordered instead of random, so that no holes remained between the adsorbed particles.
In a lattice with lateral size $L$ (in lattice units), there are $L^2/\mu$ attempts of linear
particle adsorption in the time interval $\tau$, and $L^2/\mu^2$ attempts of square particle adsorption.
The maximal adsorption rate (in an empty lattice) is $1/\tau$, in number of monolayers per second.
Due to the rejection of adsorption attempts (excluded volume effect), the actual adsorption rate
is smaller than this value and decreases in time.

The deposition time is denoted as $t$, but the model results will be presented as function of
the dimensionless time
\begin{equation}
t_D = \frac{t}{\tau} .
\label{deftD}
\end{equation}
$t_D$ is the number of incident monolayers.
Reference to real time $t$ will be usually left to discussion of possible applications.

Our model does not consider diffusion of adsorbed particles nor desorption.
This static RSA assumption is supported by works on submonolayer growth with collective
diffusion of adatoms at low temperatures \cite{tiago2012}.
The assumption may be justified at higher temperatures if the linear or square particles
represent large molecules or colloidal particles whose energy barriers for diffusion
are much larger than those of metal or semiconductor adatoms.
The energy barrier for desorption is usually larger than that for surface diffusion, thus the
no-desorption condition is also reasonable.
A recent work on electrostatic adsorption of silica nanoparticles onto Si wafers also supports
the use of irreversible RSA models \cite{li2014}.

For possible experimental tests, the model parameters can be related to the
average particle flux $F$, which is defined as the number of incident particles per unit time
and unit area.
The flux of linear particles is
\begin{equation}
F_{lin} = \frac{1}{\mu\tau a^2} .
\label{Flinear}
\end{equation}
For square particles, the flux is
\begin{equation}
F_{sq} = \frac{1}{\mu^2\tau a^2}
\label{Fsquare}
\end{equation}

% title changed
\subsection{Basic quantities and simulation procedure}
% inverted discussion

The simplest quantity to be compared with experimental data is the surface coverage $\theta$,
defined as the fraction of the surface covered with adsorbed particles.
Letting $m\left( \vec{r},t\right)$ be the occupation number of a site at position $\vec{r}$
at time $t$, with $m=1$ for occupied and $m=0$ for unoccupied, we have
\begin{equation}
\theta \left( t\right) = \langle m\left( \vec{r},t\right) \rangle ,
\label{deftheta}
\end{equation}
in which the average is taken over all $\vec{r}$ in the $xy$-plane and different realizations.

In experimental works, the surface coverage is usually given as an areal density of particles
$\rho$, which is the number of molecules adsorbed per unit area.
For linear particles, it is given by
\begin{equation}
\rho_{lin} \equiv \frac{N}{A} = \frac{\theta}{\mu a^2} ,
\label{defrholinear}
\end{equation}
and for square particles, it is given by
\begin{equation}
\rho_{sq} \equiv \frac{N}{A} = \frac{\theta}{\mu^2 a^2} .
\label{defrhosquare}
\end{equation}

Information on the adsorption dynamics can also be extracted from the size distribution of
adsorbed particles, $P\left( s,t\right)$, which is defined as the fraction
of adsorbed particles with linear size $s$ at time $t$.
If the incident flux has no dispersion ($w=0$), we have
$P\left( s,t\right)=Q\left( s\right)$;
however, in the polydisperse case, those distributions are different.
Hereafter, $P\left( s,t\right)$ is simply called adsorbed distribution.

Another important quantity is the pair correlation function
\begin{equation}
g\left( r,t\right) = \langle m\left( 0,t\right)m\left( \vec{r},t\right) \rangle - \theta^2
\qquad ,\qquad r=|\vec{r} | ,
\label{defg}
\end{equation}
in which the average is taken over different origins $0$ and different realizations; the vector
$\vec{r}$ is taken only along the $x$ and $y$ directions due to the symmetry of the square lattice.
In off-lattice RSA, $g\left( r,t\right)$ is called radial distribution function and averages are
taken along all substrate directions.
The probability of finding an occupied site at distance $r$ from another occupied site
is related to $g\left( r,t\right)$, thus this quantity measures the inhomogeneity of adsorbed
mass distribution.
The pair correlation function is advantageous over the size distributions for the applications
of RSA because it does not require the identification of the size of each adsorbed particle.

% changes in presentation of simulation details - next 2 paragraphs R1Q2
In our simulations, incident distributions with $4\leq \mu \leq 64$ and $1/16\leq w\leq 1/2$
were considered for linear and square particles.
For each set of parameters, ${10}^3$ different realizations were used to calculate
average values.

The simulation results for $w\leq 32$ presented here were obtained in lattices with lateral
size $L=4096$ (in lattice units) and periodic boundary conditions.
Some simulations in larger lattices ($L=8192$) were also performed and showed very
similar time evolution of the coverage, which indicated that finite-size effects were
negligible.
Thus, the results in lattices with $L=4096$ were representative of infinitely
large lattices with good accuracy.
For $w=64$, we present results in lattices with $L=8192$ because this was the size in
which finite-size effects became sufficiently small for linear particles.
The finite-size effects for square particle adsorption were always smaller than those for
linear particles with the same $\mu$ and $\sigma$; however, results for the same lattice sizes
are presented for both particle shapes.

\begin{figure}[!h]
\centering
\includegraphics[width=9cm]{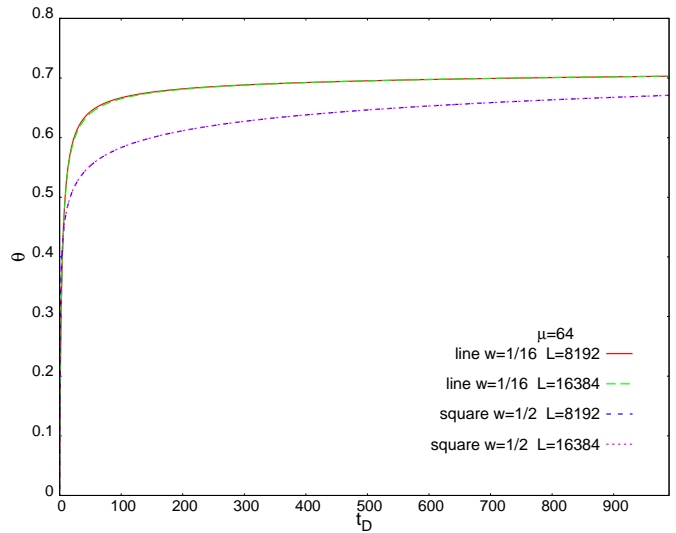}
\caption{(Color online) Check of finite-size effects by comparison of the coverage evolution
in lattices with $L=8192$ and $L=16384$ of linear particle RSA with $\mu=64$ and $w=1/16$
and square particle RSA with $\mu=64$ and $w=1/2$.
}
\label{fss}
\end{figure}

Fig. \ref{fss} shows the coverage evolution for linear and square particles with $\mu=64$
and different values of $\sigma$ in lattices with $L=8192$ and $L=16384$..
The agreement of results in these sizes indicates that the smaller one can provide
representative results for all large lattices.

\section{Adsorption of linear particles}
\label{linear}

\subsection{Adsorbate configurations and coverage evolution}
\label{linearconfigurations}

\begin{figure*}[!ht]
\centering
\includegraphics[width=12cm, trim={0 12.0cm 0 0}, clip=true]{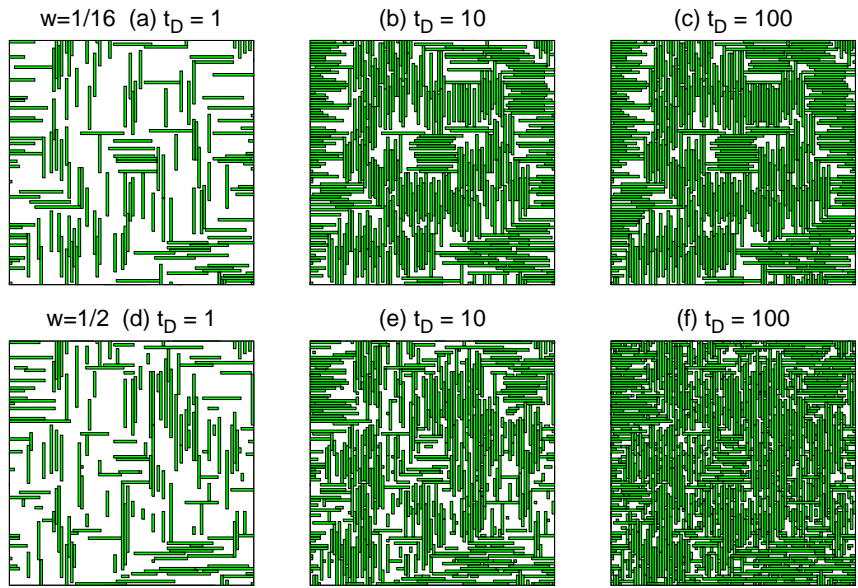}
\caption{(Color online) Time evolution of adsorbate configurations with linear particles with:
(a)-(c) $\mu =16$ and $w=1/16$; (d)-(f) $\mu =16$ and $w=1/2$.}
\label{figlinear}
\end{figure*}

Figs. \ref{figlinear}a-c show snapshots of a part of the surface at three different times
during adsorption of particles with $\mu = 16$ and $w=1/16$ ($\sigma = 1$), which is a case
of low polydispersity.
At short times (Fig.\ref{figlinear}a), spatial ordering can be observed at lengthscales
of the same order of the average particle size $\mu$, since series of particles are aligned at
neighboring rows or columns (constant $x$ or $y$).
The formation of domains of aligned particles becomes clearer as time and coverage increase
(Figs. \ref{figlinear}b,c).

\begin{figure}[!ht]
\centering
\includegraphics[width=8cm,trim={0 5.0cm 0 5.0cm}, clip=true]{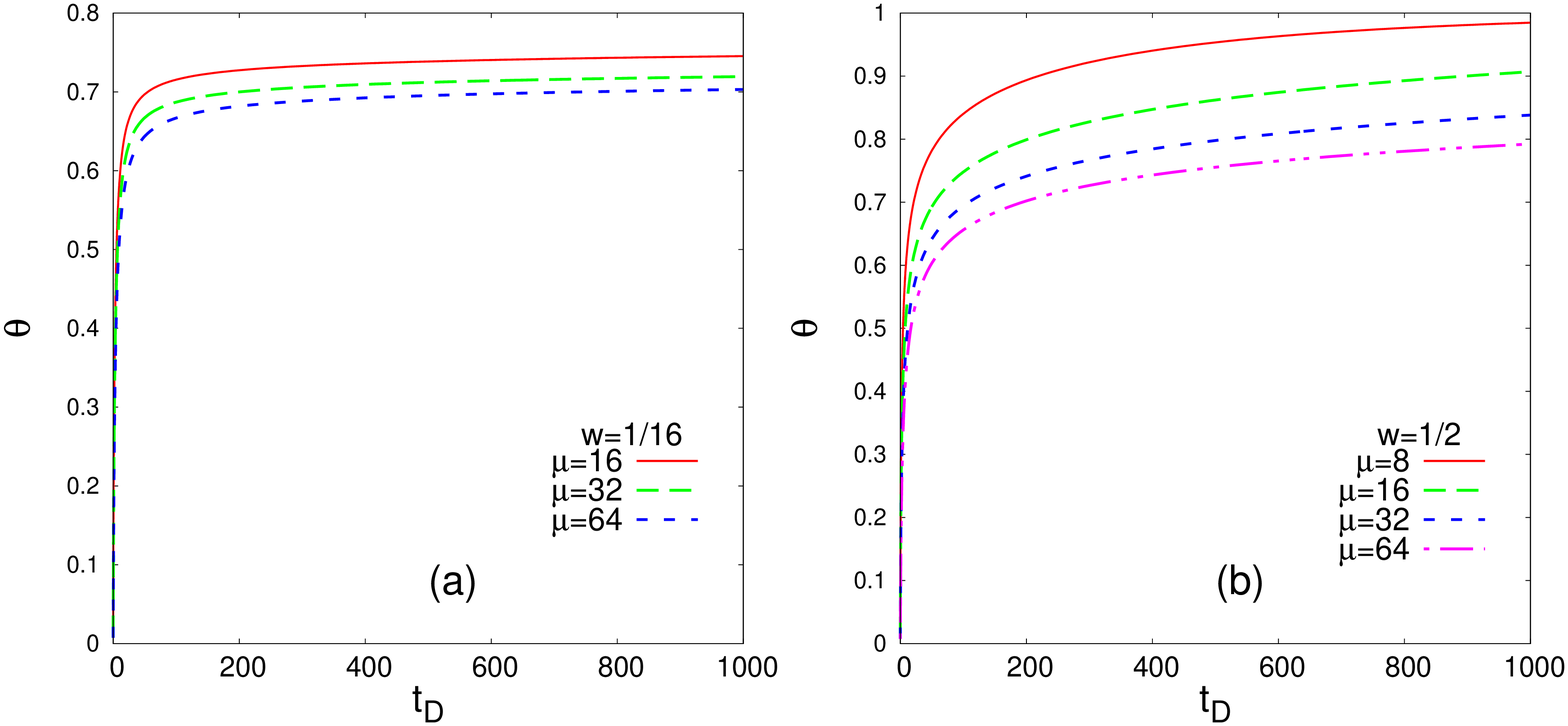}
\caption{(Color online) Coverage evolution of linear particle RSA with (a) $w=1/16$ and
(b) $w=1/2$, for several values of $\mu$.}
\label{coveragelinear}
\end{figure}

At $t_D\sim 1$, the coverage typically exceeds $20\%$, thus a particle adsorbed in a given
direction creates a large zone of exclusion for adsorption in the perpendicular direction.
On the other hand, adsorption of aligned particles is allowed in that zone, which leads to
the short range ordering.
In on-lattice RSA models, the domains of aligned particles were first observed by
Manna and Svrakic \cite{manna} and were recently illustrated in Ref. \protect\cite{lebovka2015}.
They were also shown in studies of anisotropy effects \cite{budinski2011,lebovka2011}
and in off-lattice RSA \cite{ziff,viot,cieslaPRE2013}.
These configurations resemble the nematic ordering observed for high densities in
thermodynamic equilibrium adsorption of linear particles
\cite{onsager1949,ghosh2007,dhar2011,kundu2013,kundu2014}.
However, no long-range order and no phase transition is present in the irreversible RSA models.

Figs. \ref{figlinear}d-f show snapshots of a part of the surface during adsorption of particles
with $\mu = 16$ and $w=1/2$ ($\sigma = 8$), which is a case of high polydispersity.
Due to the enhanced flux of small particles, the aligned particle domains are smaller.
As time increases, this short-range ordering is not enhanced.

Figs. \ref{coveragelinear}a and \ref{coveragelinear}b show the time evolution of the coverage
for $w=1/16$ (low polydispersity) and $w=1/2$ (high polydispersity), respectively, with
several average sizes $\mu$ in each case.
These plots show the typical downward curvature of irreversible RSA problems because the
fraction of the surface available for adsorption of new particles decreases as $t$ (and
$\theta$) increases.

The asymptotic coverages are $\theta_\infty =1$ because the monomer flux in nonzero
in all distributions, which will eventually leads to complete filling.
However, at the maximum simulated time $t_D=1000$, the coverage is near $1$ only for the smallest
$\mu$ ($=8$) and the largest $w$ ($=1/2$) because this condition provides a high monomer flux.

\subsection{Time scaling of the coverage}
\label{linearcoverage}

\begin{figure*}[!ht]
\centering
\includegraphics[width=14cm]{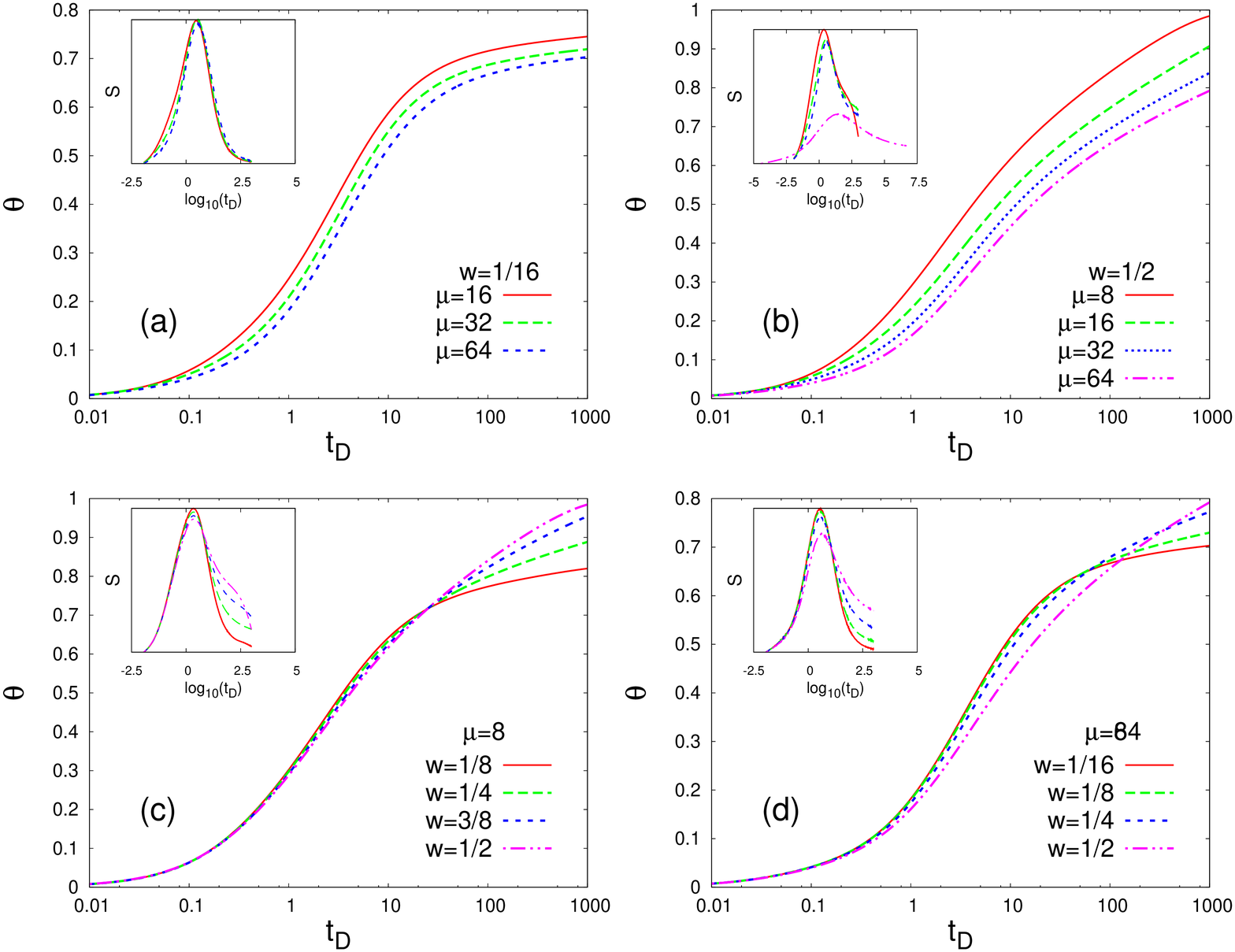}
\caption{(Color online) Coverage evolution of linear particle RSA in log-linear scale with fixed
$w$ and variable $\mu$ [(a), (b)] and fixed $\mu$ and variable $w$ [(c), (d)].
The insets show the time evolution of the slope $S$ of the main plots.}
\label{covlinearlog}
\end{figure*}

For $t_D\ll 1$, the coverage is very small, thus almost all adsorption attempts are accepted.
For $t_D\sim 1$, the coverage reaches values between $0.1$ and $1$; in this case, the increase of
coverage is slower due to the rejection of many adsorption attempts.
For $t_D\gg 1$, the coverage is large (of order $1$), thus adsorption of new particles is rare
and $\theta$ increases very slowly.
The linear plots of Figs. \ref{coveragelinear}a-b highlight the third regime; the first two
regimes are hidden in a very narrow region near the vertical axis.

% removed comment at the end of paragraph R2Q1
The exactly solvable problem of RSA of monomers show the same regimes \cite{privman}.
The coverage in monomer RSA is $\theta\left( t_D\right) = 1-\exp{\left( -t_D\right)}$.
In that case, plots of $\theta$ as function of $\log{\left( t_D\right)}$ are more helpful to
distinguish the different scaling regimes discussed above.
The slope of a $\theta\times \log{\left( t_D\right)}$ plot is
\begin{equation}
S\equiv \frac{d\theta}{d\left( \log{t_D}\right)} = t_D \frac{d\theta}{d t_D} .
\label{defS}
\end{equation}
In RSA of monomers, we obtain $S=t_D\exp{\left( -t_D\right)}$, which has a peak at $t_D=1$
($t=\tau$).
Thus, the largest slope of that plot may be used to estimate $\tau$, which in turn is related
to the molecular flux.

Figs. \ref{covlinearlog}a,b show the same data of Figs. \ref{coveragelinear}a,b for linear
particle RSA with the abscissa replaced by $\log{\left( t_D\right)}$.
Figs. \ref{covlinearlog}c and \ref{covlinearlog}d show $\theta$ as function of 
$\log{\left( t_D\right)}$ for $\mu =8$ and $\mu=64$, respectively, with four different values
of $w$ in each case.
The time evolution of $S$ is shown in the insets of Figs. \ref{covlinearlog}a-d for the
same parameters of the main plots.

The dimensionless time $t_S$ is defined as that of the maximal value of $S$.
The insets of Figs. \ref{covlinearlog}a-d confirm the conclusion that $t_S$ is always of order $1$,
similarly to the RSA of monomers.
A careful inspection of the peaks of $S$ gives
\begin{equation}
0.2<\log{\left( t_S\right)}<0.7
\label{tSlinear}
\end{equation}
for all values of $\mu$ and $w$.

The corresponding real times of the peaks of $S$ are denoted as $t_M\equiv \tau t_S$;
they are approximately in the range $1.5\tau<t_M<5\tau$.
The extreme values of this range are characteristic of the broadest distributions
($w=1/2$).
For fixed $\mu$, the increase of $w$ leads to a small shift of $t_S$ to larger values;
for fixed $w$, the increase of $\mu$ also leads to that shift.

The range of the coverage in which the peak of $S$ is observed
is between $\theta_{min}\approx 0.2$ and $\theta_{max}\approx 0.5$.
In most cases, the coverage at $t_S$ is near $\theta_S\sim 0.3$.

The universal location of the peaks of the slope $S$ is a remarkable feature of
these RSA models and may be explored in applications.
For instance, consider an adsorption process in which the peak of 
$d\theta/d\left( \log{t}\right)$ is measured at (real) time $t=t_M=\tau t_S$.
Eq. \ref{tSlinear} may be inverted to give an estimate of $\tau$ as
\begin{equation}
0.2 t_M < \tau < 0.7 t_M .
\label{tMtau}
\end{equation}
Now $\tau$ can be related to the molecular flux and to the sticking coefficient by
Eq. (\ref{Flinear}).
If the average size $\mu$ and the lattice constant $a$ are also known, then Eq. (\ref{Flinear})
gives an estimate of $F_{lin}$.

If this estimate differs from the value of $F_{lin}$ predicted by the properties of the
surrounding gas or solution, a possible explanation is the existence of a sticking coefficient
$c<1$.
This coefficient is the probability that the adsorption of the incident particle actually occurs
when it is not forbidden by the excluded volume condition; in our simulations, $c=1$ was assumed.

% corrected 0.4->0.3 R2Q2
The estimate $\theta_S\sim 0.3$ can be used with Eq. \ref{defrholinear} to find an order
of magnitude of the adsorbed mass.
From Eq. \ref{defrholinear}, the total number of adsorbed particles in a substrate area
$A$ is $N\sim 0.3 A/\left( \mu a^2\right)$.
Thus, if the total adsorbed mass is $M$ and the average particle mass is $M_P$, we obtain
$M\sim 0.3 M_PA/\left( \mu a^2\right)$ at $t=t_M$.

Note that these results are valid for any average particle size $\mu$ and relative width $w$.
Although they do not predict accurate values, the knowledge of orders of magnitude of the
quantities involved may be a first step to improve experimental work or modeling of a given
process.

\subsection{Extrapolation of the surface coverage}
\label{linearcoveragemore}

% equation corrected to time t_D
The convergence to the asymptotic coverage $\theta_\infty =1$ is expected to be exponential,
similarly to other irreversible RSA problems on lattices \cite{privman,budinski2008,cornette2007}:
\begin{equation}
\theta = 1 -C\exp{\left( -t_D/t_R \right)} ,
\label{thetainf}
\end{equation}
where $C$ and $t_R$ are constants. 

\begin{figure}[!h]
\centering
\includegraphics[width=8cm, trim={0 4.0cm 0 3.0cm}, clip=true]{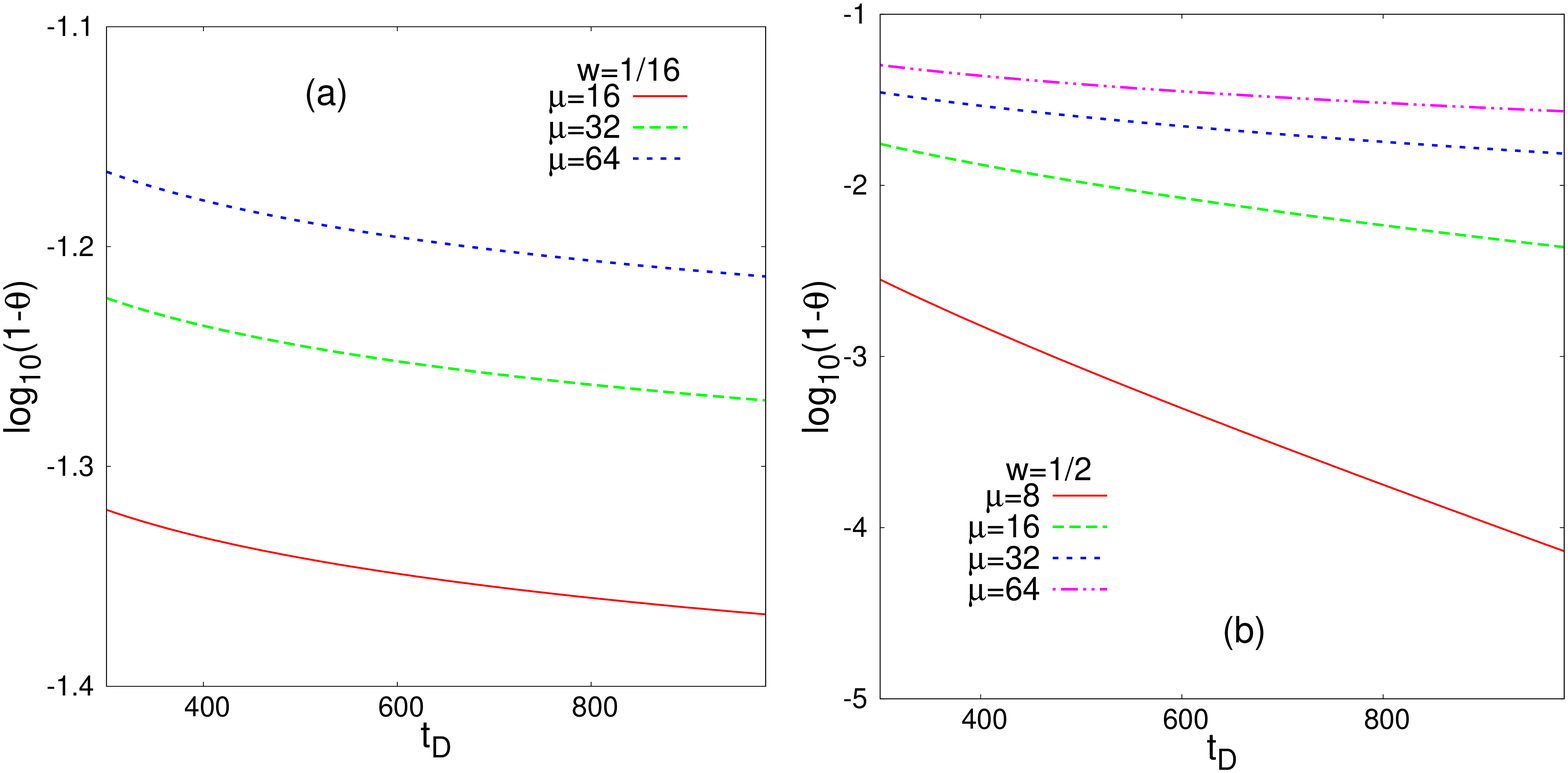}
\caption{(Color online) Long time evolution of the coverage for (a) $w=1/16$ and (b) $w=1/2$
for several values of $\mu$.}
\label{asymptoticlinear}
\end{figure}

% paragraph split with many changes R2Q3, R2Q4
Fig. \ref{asymptoticlinear} shows $\log{\left( 1 -\theta\right)}$ as function of
$t_D$ for some values of $\mu$ and $w$.
An approximately linear decrease of $\log{\left( 1 -\theta\right)}$
is observed at the longest times, typically starting at $t_D$ between $500$ and $700$.
We estimated the parameters $t_R$ and $C$ in Eq. (\ref{thetainf}) from linear fits
and show the values of $t_R$ in Table \ref{trlinear};
the values of $C$ (not shown) are in the range $\left[ 0.276,0.416\right]$.

\begin{table}[!h]
\centering

\label{trlinear}
\begin{tabular}{|r|c|c|c|c|}
\hline
$\mu \backslash  w$ & 1/16  & 1/8   & 1/4  & 1/2  \\ \hline
8           &       & 9530 & 2790 & 450 \\ \hline
16          & 22700 & 10500 & 4150 & 1270 \\ \hline
32          & 23700 & 11100 & 5050 & 2260 \\ \hline
64          & 23600 & 11400 & 5500 & 3090 \\ \hline
\end{tabular}
\caption{Relaxation time $t_R$ of linear particle RSA.}
\end{table}

These data clearly show that the relaxation time $t_R$ increases as $w$ decreases and $\mu$
increases.
The constant $C$ has the same dependence on $w$ and $\mu$, but with much smaller variation.
In all cases, a faster relaxation is expected for a larger flux of monomers
and other small particles, since they are able to fill the narrow gaps between the previously
adsorbed particles.
Also note that $t_R$ is much larger than the maximal simulated time $t_D=1000$ in most cases;
this is consistent with the large differences from full coverage in most of our data.
However, this also means that the uncertainties in the estimates in Table \ref{trlinear} become 
larger; this is the probable reason why the trend of $t_R$ increasing with $\mu$ apparently
fails for the largest particle sizes with $w=1/16$.

Better fits of $1-\theta$ may be obtained with stretched exponential decays similar to
those proposed in Refs. \protect\cite{loncarevic2010,dujac}.
However, the present problem fits the theoretical approach of Ref. \protect\cite{cornette2007},
which supports the simple exponential form in Eq. (\ref{thetainf}).
For this reason, we understand that the deviations from the linear behavior in
Fig. \ref{asymptoticlinear} are indicative of scaling corrections to Eq. (\ref{thetainf}).
Such corrections may help to explain the deviations from the expected trends in the
estimates of long times $t_R$ discussed above.

For fixed $w$, the average size $\mu$ has a significant effect on the coverage at
$t_D\sim 1$ or longer, as shown in Figs. \ref{covlinearlog}a,b.
At $t_D\gg 1$, the general trend is that $\theta$ decreases as $\mu$ increases.
This effect is more pronounced for broad distributions (Figs. \ref{covlinearlog}b; $w=1/2$).

For fixed $t_D$ and $w$, $\lim_{\mu\to\infty}{\theta}=\theta_n$ is finite, where $\theta_n$ is
the coverage of infinitely long needles. This limit is not equivalent to off-lattice
aggregation of needles because here the adsorption is restricted to the $x$ and $y$ directions.
In order to estimate $\theta_n\left( t\right)$, we assume that
\begin{equation}
\theta = \theta_n +C' \mu^{-\alpha} 
\label{thetan}
\end{equation}
for fixed $w$ and $t_D$.

\begin{figure*}[!ht]
\centering
\includegraphics[width=12cm, trim={0 0.0cm 0 0.0cm}, clip=true]{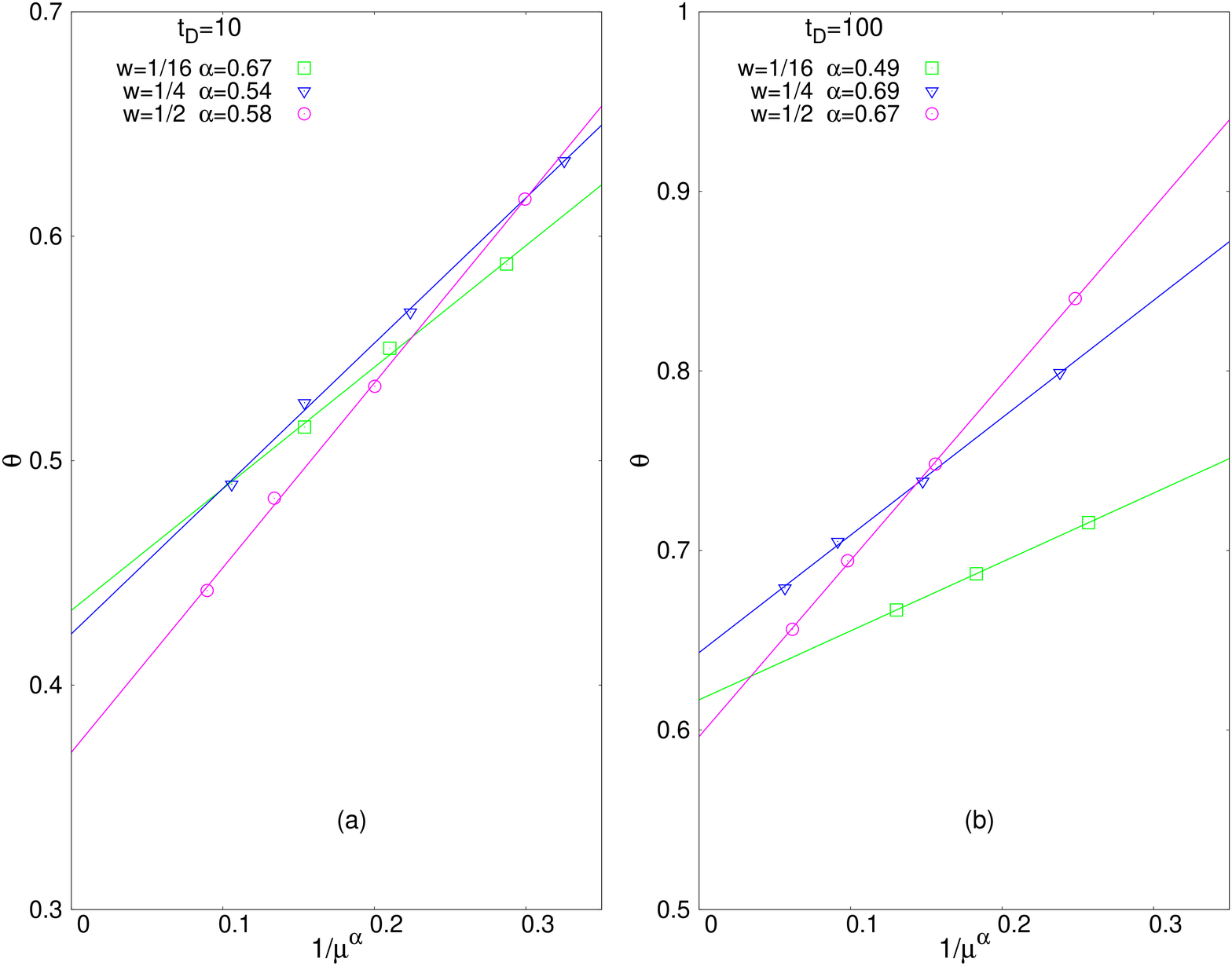}
\caption{(Color online) Extrapolation of coverage as function of particle size at fixed $t_D$
and $w$ according to Eq. (\ref{thetan}). The fitting exponents $\alpha$ are shown in Table
\ref{tablelinear}.}
\label{extraplinear}
\end{figure*}

In Figs. \ref{extraplinear}a and  \ref{extraplinear}b, we show $\theta$ as function of
$\mu^{-\alpha}$ for different values of $w$ and $t_D$, with exponents $\alpha$ chosen to
provide the best linear fits of each data set.
Those fits give the coverage $\theta_n$ in the limit $\mu\to\infty$ ($\mu^{-\alpha}\to 0$).
Table \ref{tablelinear} shows the estimates of $\theta_n$ and the fitting exponents
$\alpha$ for various $w$ and times $t_D=10$ and $t_D=100$.

\begin{table}[!h]
\centering

\label{tablelinear}
\begin{tabular}{|r|c|c|c|c|c|c|}
\hline
$t_D$                        & \multicolumn{3}{c|}{10} & \multicolumn{3}{c|}{100} \\ \hline
$w$                          & 1/2   & 1/4   & 1/16   & 1/2    & 1/4   & 1/16   \\ \hline
$\theta_n$                   & 0.3698 & 0.4227 & 0.4330 & 0.5961 & 0.6432 & 0.6165    \\ \hline
$\alpha$                   & 0.58 & 0.54 & 0.45 & 0.67 & 0.69 & 0.49    \\ \hline
$\theta\left(\mu=64\right)$ &  0.4419 & 0.4894 & 0.5147 & 0.6559 & 0.6792 & 0.6667 \\ \hline
$\Delta\theta$ (\%)          &  16.3 & 13.6 & 15.9 & 9.1 & 5.3 & 7.5  \\ \hline  
\end{tabular}
\caption{Estimated coverage $\theta_n$ of infinitely long lattice needles, fitting exponents $\alpha$,
coverage $\theta\left(\mu=64\right)$ of the largest simulated linear particles, and relative difference
$\Delta\theta$ between those coverages, for the indicated distribution widths $w$ and times $t_D$.}
\end{table}

% paragraphs split and changed - R2Q5
The coverage for the maximal simulated size $\mu=64$ is also shown in Table \ref{tablelinear}
for comparison with $\theta_n$.
The relative difference between them ranges from $5\%$ to $20\%$.
Since the above extrapolations considered a restricted range of values of $\mu$, the estimates
of $\theta_n$ may also have a large uncertainty.
Thus, the coverage for the largest simulated size, $\mu=64$, may be very different
from that of larger $\mu$.

On the other hand, in the region of the peaks of $S$ (Fig. \ref{covlinearlog}), the changes
in the coverage are much larger than $20\%$.
Thus, the value of $\log{t_S}$ in Eq. \ref{tSlinear} (and consequently the order of magnitude
of $t_S$) is not expected to have significant change for other values of $\mu$ and $w$.

\subsection{Adsorbed particle size distributions}
\label{lineardistribution}

\begin{figure}[!ht]
\centering
\includegraphics[width=8cm]{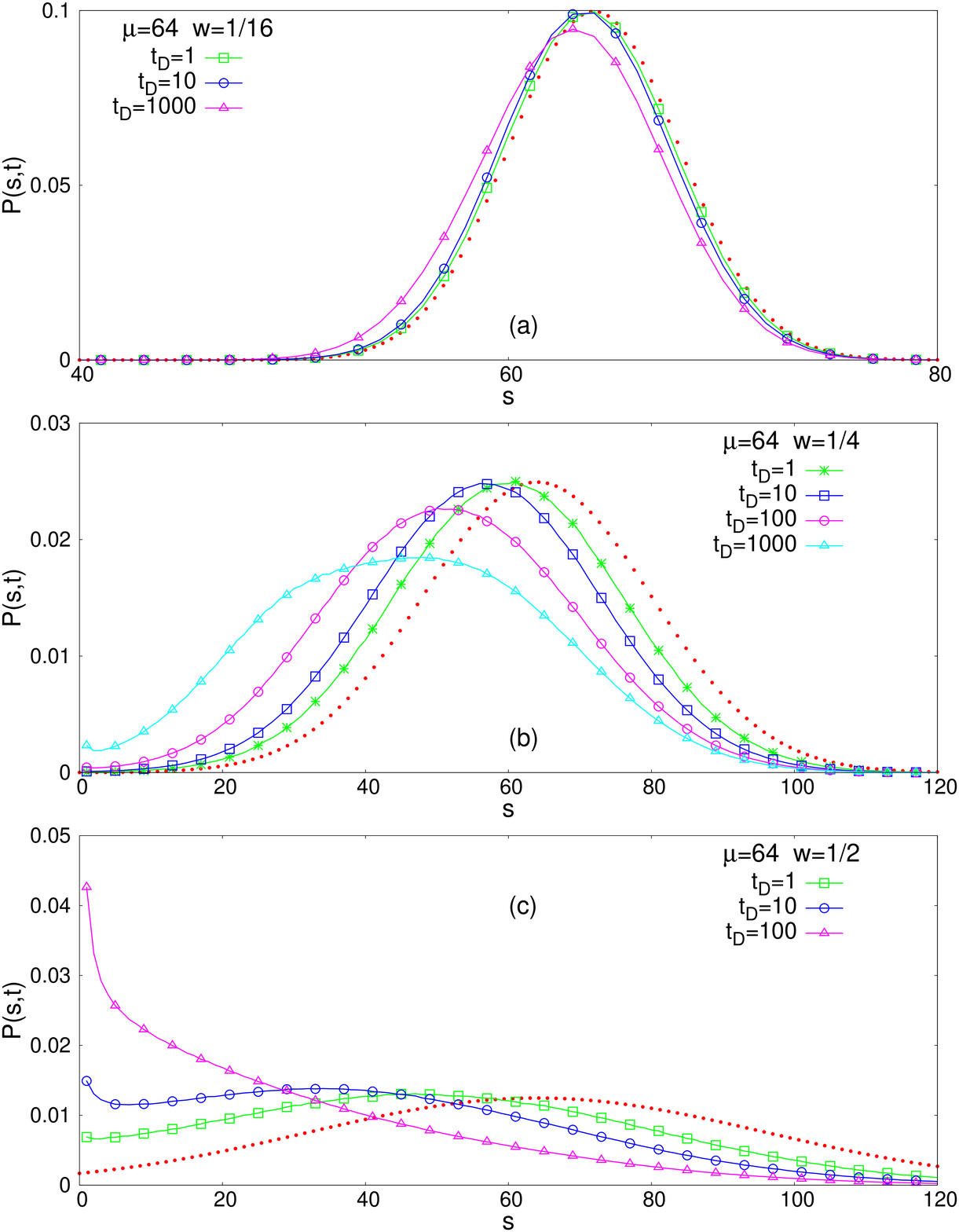}
\caption{(Color online)
Solid curves are adsorbed distributions of linear particles with incident average size
$\mu=64$ and several values of width $w$ and time $t_D$. The dotted curves are the incident
distributions $Q\left( s\right)$.}
\label{distrlinear}
\end{figure}

Figs. \ref{distrlinear}a-c show the adsorbed distributions $P\left( s,t\right)$
as function of the particle size $s$ at several times for $\mu =64$ and three values of $w$.

If the incident distribution has small width ($w=1/16$; Fig. \ref{distrlinear}a),
the difference from the adsorbed distribution is small up to long times.
The coverage at $t_D=1000$ is $\theta\approx 0.7$.
Possibly there will be significant changes in $P\left( s,t\right)$ when $\theta$
is close to $1$, but this regime is expected to be observed at times longer by several
orders of magnitude.

For an intermediate distribution width ($w=1/4$; Fig. \ref{distrlinear}b), a small shift of
the adsorbed distribution peak to $s<\mu$ is observed at $t_D=1$; at $t_D\approx 100$,
broadening of the distribution begins, still accompanied by the shift of the peak to smaller $s$.
For the largest width ($w=1/2$; Fig. \ref{distrlinear}c), a significant shift of the distribution
peak is observed at $t_D\approx 1$.
Subsequently, a secondary peak at $s=1$ (monomers) appears and the distribution acquires a
monotonically decreasing shape at $t_D\geq 100$.

The shift of the peak to smaller sizes and the broadening of the distribution (here observed
for $w\geq 1/4$) parallel the features reported by Meakin and Jullien \cite{meakin} for RSA
of disks and by Adamczyk et al \cite{adamczyk1997} for RSA of spheres with Gaussian distributions
of incident sizes.
Recently, Marques et al \cite{marques} studied the RSA of polydisperse disks on patterned
substrates, showing adsorbed distributions with a secondary peak in small $s$ at short times
and a monotonically decreasing shape at long times.
This is similar to our findings for $w=1/2$.
However, in lattice RSA, the monotonic decay was formerly observed only with uniform incident
distributions \cite{hanarp,danwanichakul,budinski2008} and was independent of the distribution
width and deposition time.

Size distributions with a peak at small sizes or showing a monotonic decrease are
frequently observed in temperature-driven coarsening.
This is illustrated, for instance, in one-dimensional island coarsening without deposition
\cite{robinreis2008} and in submonolayer growth dominated by surface diffusion \cite{submonorev}.
In general, high temperature favors those features, thus they may be interpreted as disorder
effects.
In the athermal RSA models, disorder is represented by the width $w$ of the incident
distribution.

Since the peak of the coverage derivative $S$ is located at the universal region $1.5<t_S<5$
(Eq. \ref{tSlinear}), it is interesting to investigate the adsorbed distributions in this
time interval.
For $w\leq 1/8$, that distribution is still very close to the incident one.
For $w=1/4$, $P\left( s,t_S\right)$ has an approximately Gaussian shape, but with a small
shift of the peak to $s<\mu$ (Fig. \ref{distrlinear}b).
For $w=1/2$, significant broadening and formation of a peak at small $s$ is observed
in that time range.
In real adsorption processes, incident and adsorbed distributions at $t=t_M$ may be compared
and used to check the applicability of this model.
However, estimating $P\left( s,t\right)$ may be a difficult task because
it is necessary to measure the individual sizes of tightly packed particles.
Moreover, the size distribution statistics is frequently poor due to the small number of samples
and small image sizes.

% trends for much longer times - R1Q3
Our simulations are limited to $t_D\leq 1000$, but we may speculate about the distributions at
much longer times.
The jamming coverages $\theta_{jam}{\left(\mu\right)}$ for RSA of linear particles of fixed size
$\mu$ \cite{bonnier} are helpful for this discussion: $\theta_{jam}{\left( 2\right)}\approx 0.91$,
$\theta_{jam}{\left( 8\right)}\approx 0.75$,
$\theta_{jam}{\left( 16\right)}\approx 0.71$, $\theta_{jam}{\left( 32\right)}\approx 0.689$,
$\theta_{jam}{\left( 64\right)}\approx 0.68$.
Inspection of Figs. \ref{coveragelinear} shows that all these values have been exceeded in
RSA of our polydisperse mixtures at $t_D= 1000$, thus it is very difficult that particles with size
$s\approx \mu$ can adsorb at longer times.
Adsorption at longer times is consequently dominated by small particles, with $s\ll \mu$,
which will cover the remaining empty sites ($20\%$ to $30\%$). 
Since the jamming coverage for dimers ($\mu=2$) is $\sim 10\%$ below the full coverage,
we expect that full coverage occurs with a monomer fraction of this order.

In cases of broad incident distributions, e. g. $w=1/2$, we expect that the monotonic decay shown
in Fig. \ref{distrlinear}c is enhanced at much longer times.
In cases of narrow incident distributions, e. g. $w=1/16$, the densities of large adsorbed particles
at $t_D=1000$ is large ($P\approx 0.1$ and half-width $\Delta s\approx 10$).
Thus, we expect that the peak of the adsorbed distribution in Fig. \ref{distrlinear}a is slowly shifted
to smaller $s$ as time increases, but it is not expected to disappear.
Instead, near full coverage, this peak is expected to exist
together with a monotonically decreasing region beginning at $s=1$.

\subsection{Pair correlation function}
\label{linearcorrelation}

\begin{figure}[!h]
\centering
\includegraphics[width=8cm]{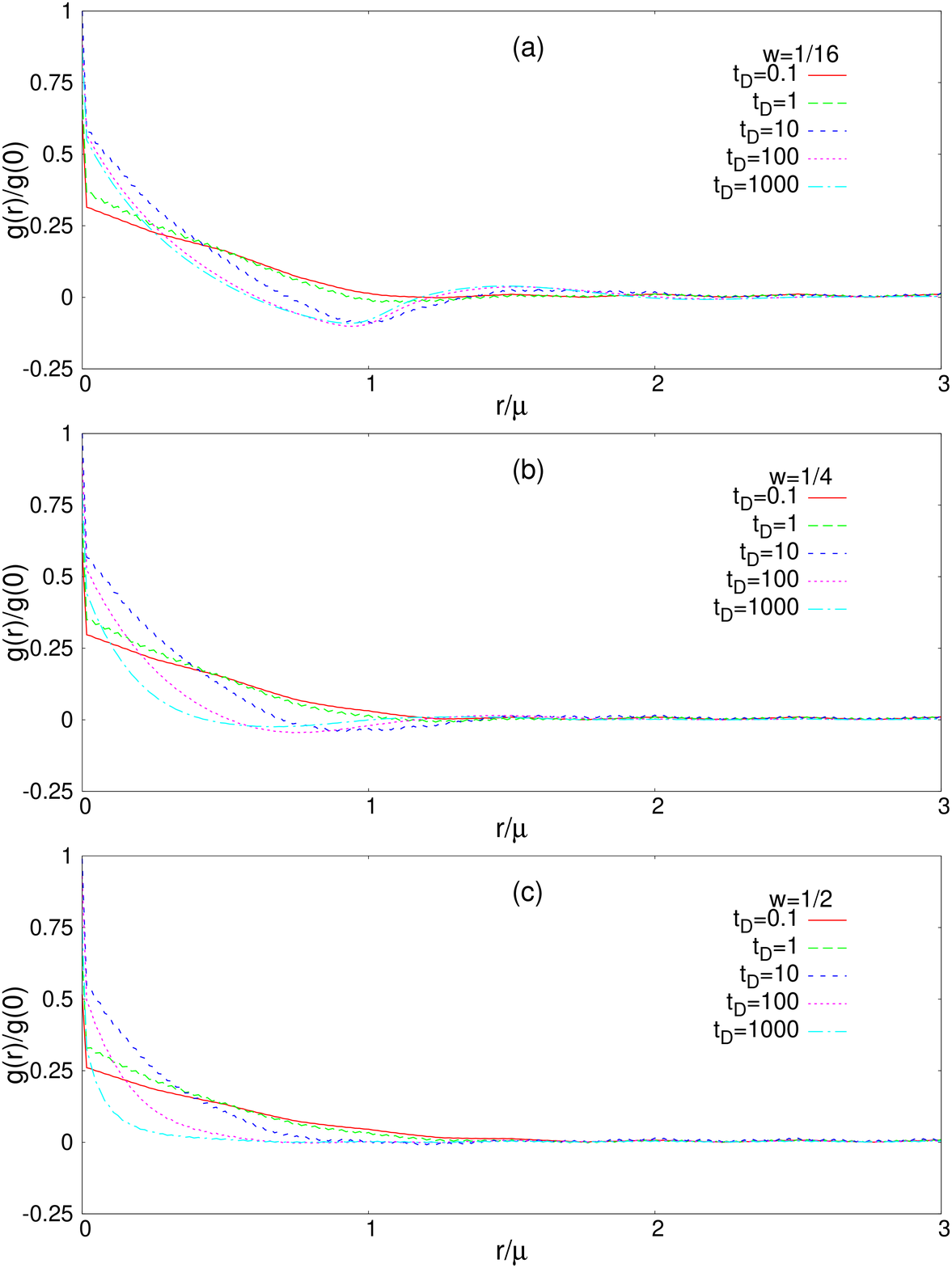}
\caption{(Color online) Scaled pair correlation as function of the scaled distance for linear
particles with average incident size $\mu=64$ and several values of width $w$ and time $t_D$.}
\label{glinear}
\end{figure}

Figs. \ref{glinear}a-c show $g\left( r,t\right)/g\left( 0,t\right)$ as function of the scaled
distance $r/\mu$ for several times and several values of $\mu$ and $w$.
The results for $\mu =8$ are quantitatively similar, thus the results in Figs. \ref{glinear}a-c
are representative of linear particles of all lengths.
Those plots highlight features for $r$ on the same order of magnitude of $\mu$, similarly to
works on models with continuous size distributions \cite{adamczykbook}.

\begin{figure*}[!ht]
\centering
\includegraphics[width=12cm, trim={0 11.0cm 0 0.0cm}, clip=true]{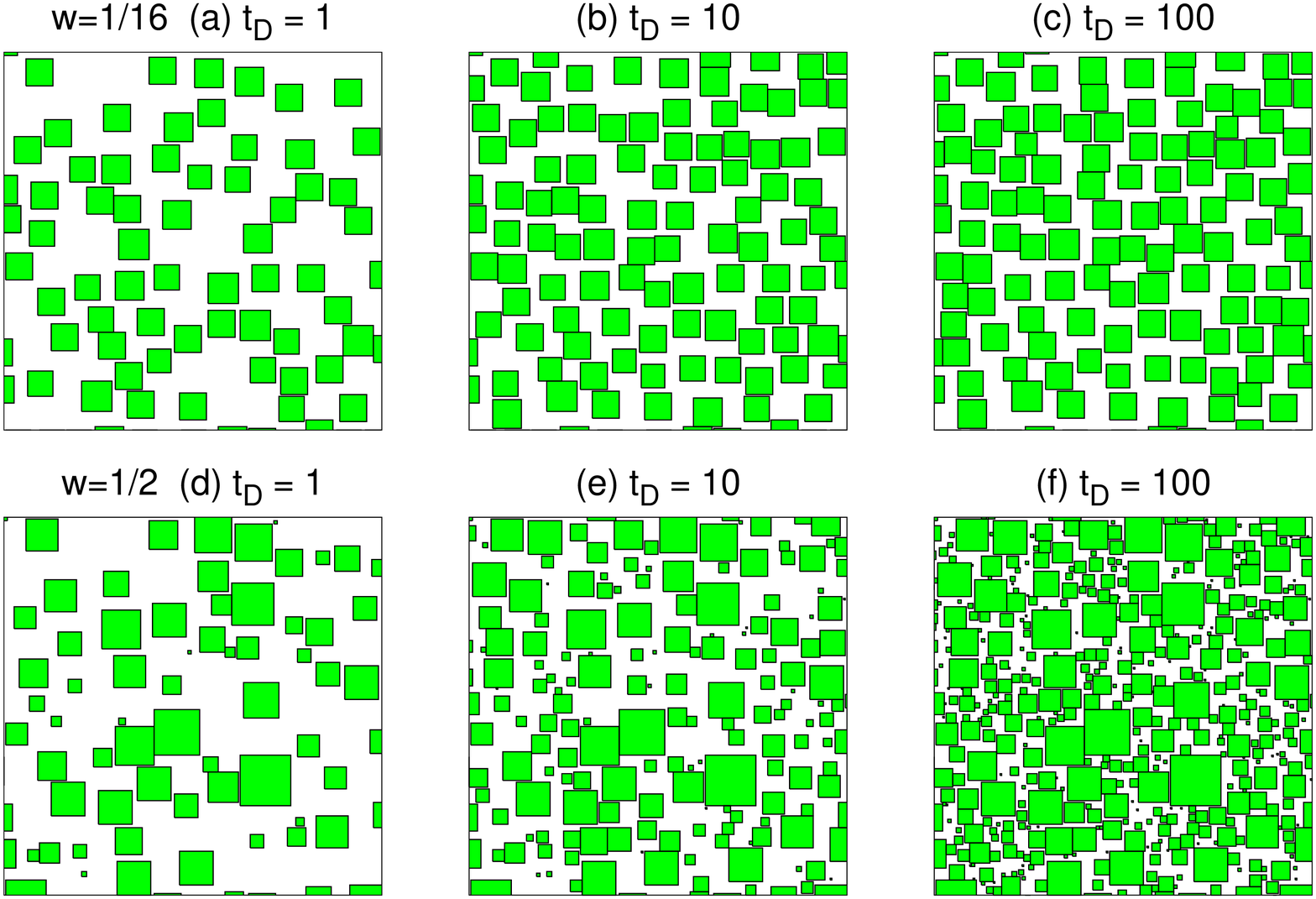}
\caption{(Color online) Time evolution of adsorbate configurations with square particles with:
(a)-(c) $\mu =16$ and $w=1/16$; (d)-(f) $\mu =16$ and $w=1/2$.}
\label{figsquare}
\end{figure*}

For $w=1/16$ (Fig. \ref{glinear}a), $g$ monotonically decreases at short times
and acquires a minimum at $r\approx \mu$ at $t_D\approx 10$.
This minimum remains at longer times, up to coverages near $70\%$ that are obtained
at $t_D=1000$.
The typical size of blocks of ordered particles is close to the average length $\mu$,
as illustrated in Figs. \ref{figlinear}a-c.
This is the typical distance between regions of filled and empty sites, thus the minimum of
$g\left( r\right)$ is a consequence of the ordering of adsorbed particles.
The typical distance between the first mininum of $g$ and
the subsequent maximum is $\Delta r\sim 0.5\mu$ because the depleted regions range from one
lattice site to the average incident size $\mu$.
Similar results are observed in adsorption of monodisperse particles \cite{lebovka2011}.

For $w=1/4$ (Fig. \ref{glinear}b), the correlation function has a very shallow minimum at
$t_D\approx 10$.
The shift of this minimum to smaller $r/\mu$ is significant as time increases, but the depth is
still very small.
For $w=1/2$, shallow minima of $g$ with faster displacement to smaller $r/\mu$ are observed.

In the time range of $t_S$ (Eq. \ref{tSlinear}), the minimum of $g$ is beginning to be formed
if $w\leq 1/8$.
Inspection of g at $t_D=10t_S$ is necessary to confirm the presence of this
minimum.
This feature may be used to identify a narrow incident distribution, with an upper bound
for the width close to $1/8$.
For $w\geq 1/4$, it is very difficult to identify some oscillatory structure in $g$
at times of order $10t_S$ or longer.
In this case, this flat shape of $g\left( r,t_S\right)$ gives a lower bound $\approx 1/4$
for the width of the incident size distribution.

The study of the pair correlation function is suitable for this type of investigation
because it does not require identification of sizes of individual adsorbed particles,
in contrast to the comparisons of size distributions.
$g\left( r,t\right)$ is also advantageous over the coverage because it probes spatial
organization.

\section{Adsorption of square particles}
\label{square}

\subsection{Adsorbate configurations}
\label{squareconfigurations}

Figs. \ref{figsquare}a-c show snapshots of a part of the surface at three different times
during adsorption of particles with $\mu = 16$ and $w=1/16$ ($\sigma = 1$), which is a case
of low polydispersity.
The short-range structure differs from that of linear particles because there is no alignment.
However, the inhomogeneity of mass distribution is similar: there are depleted regions
with lateral sizes between $1$ and $\mu$ in which particles with the average size cannot adsorb.

Figs. \ref{figsquare}d-f show snapshots of a part of the surface at three different times
during adsorption of particles with $\mu = 16$ and $w=1/2$ ($\sigma = 8$), which is a case
of high polydispersity.
The much higher flux of small particles fills the gaps between particles of average size,
thus the adsorbate is more homogeneous, particularly for the longer times.

\subsection{Coverage evolution}
\label{squarecoverage}

\begin{figure*}[!ht]
\centering
\includegraphics[width=14cm]{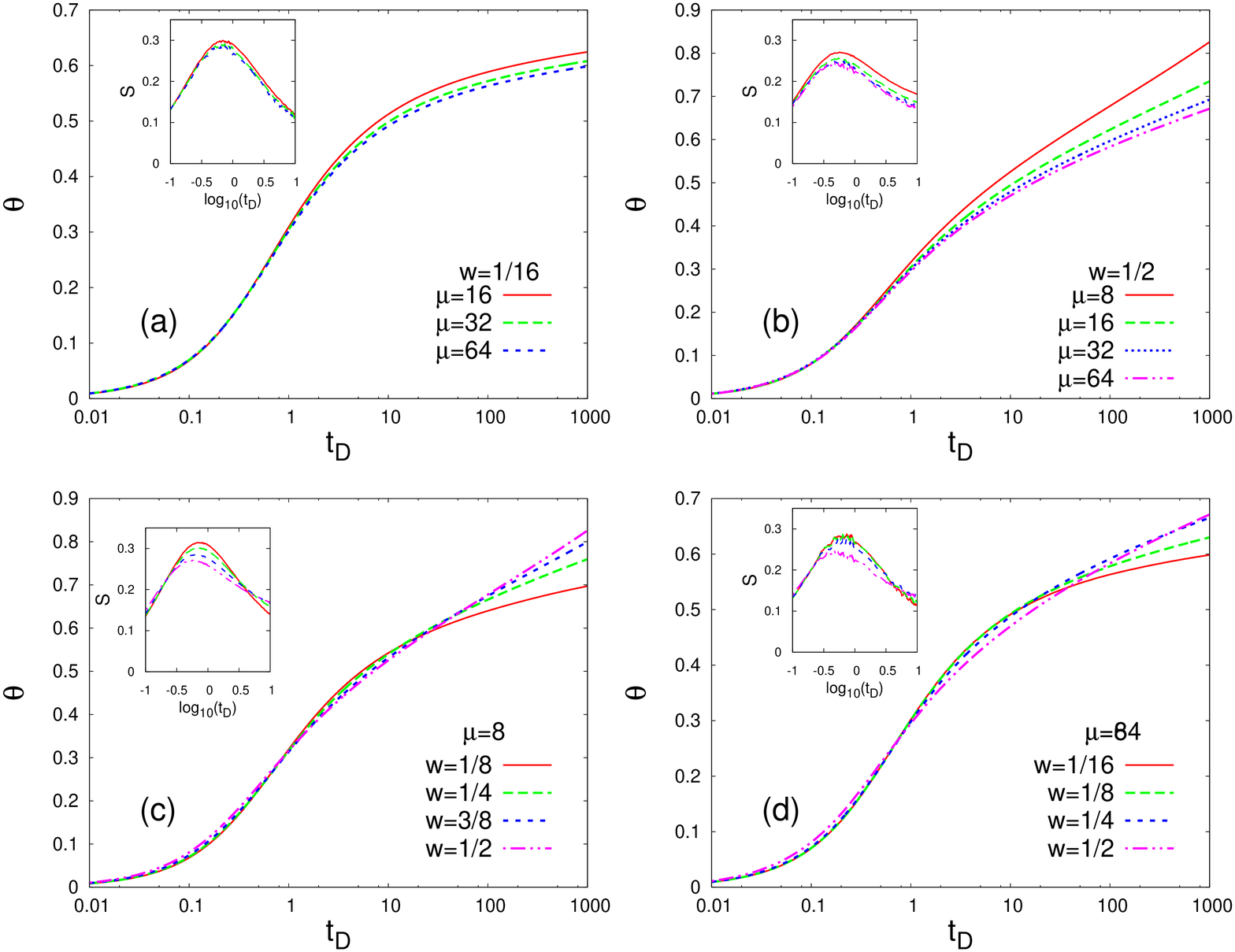}
\caption{(Color online) Coverage evolution of square particle RSA in log-linear scale with fixed
$w$ and variable $\mu$ [(a), (b)] and fixed $\mu$ and variable $w$ [(c), (d)].
The insets show the time evolution of the slope $S$ of the main plots.}
\label{coveragesquare}
\end{figure*}

Figs. \ref{coveragesquare}a and \ref{coveragesquare}b show the time evolution of the coverage
for $w=1/16$ and $w=1/2$, respectively, with various average sizes $\mu$ in each case.
Figs. \ref{coveragesquare}c and \ref{coveragesquare}d show the time evolution of $\theta$
for $\mu=8$ and $\mu=64$, respectively, with various widths $w$ in each case.
The plots with logarithmic time scale also show regions of maximal slopes for
$t_D\sim 1$, similarly to the adsorption of linear particles.
This is confirmed in the insets of Figs. \ref{coveragesquare}a-d, which show the evolution
of the slope $S$.

If $w$ is fixed and $\mu$ increases, a small shift of $t_S$ to shorter times is
observed, in contrast with the linear case.
The same shift is observed if $\mu$ is fixed and $w$ increases.
However, the $S$ peaks are also located in a narrow range of $t_D$ in logarithmic scale;
for all values of $\mu$ and $w$ studied here, we obtain
\begin{equation}
-0.5<\log{\left( t_S\right)}<0 .
\label{tSsquare}
\end{equation}
The coverages in this time interval range from $\approx 0.1$ to $\approx 0.3$, thus the
typical value of the coverage at $t_S$ is of order $\theta_S\sim 0.2$.

If this model is applied to an adsorption process in which the peak of
$d\left( \theta\right)/d\left( \log{t}\right)$ is measured at real time $t_M\equiv \tau t_S$,
then Eq. (\ref{tSsquare}) leads to $0.3\tau < t_M < \tau$.
This relation can be inverted to give the estimate
\begin{equation}
t_M < \tau < 3t_M .
\label{tPtausquare}
\end{equation}
The time $\tau$ can be related to the molecular flux by Eq. (\ref{Fsquare});
if this flux is known, then a sticking coefficient may be estimated.

The above estimate of $\theta_S$ and Eq. \ref{defrhosquare} give
the total number of adsorbed particles in a substrate area $A$ as
$N\sim 0.2 A/{\left( \mu a\right)}^2$ at $t=t_M$.
If the total adsorbed mass is $M$ and the
average particle mass is $M_P$, we obtain $M\sim 0.2 M_PA/{\left( \mu a\right)}^2$.
These results are valid for any $\mu$ and $w$, but again they relate only the orders of
magnitude of those quantities.

\begin{figure}[!h]
\centering
\includegraphics[width=8cm]{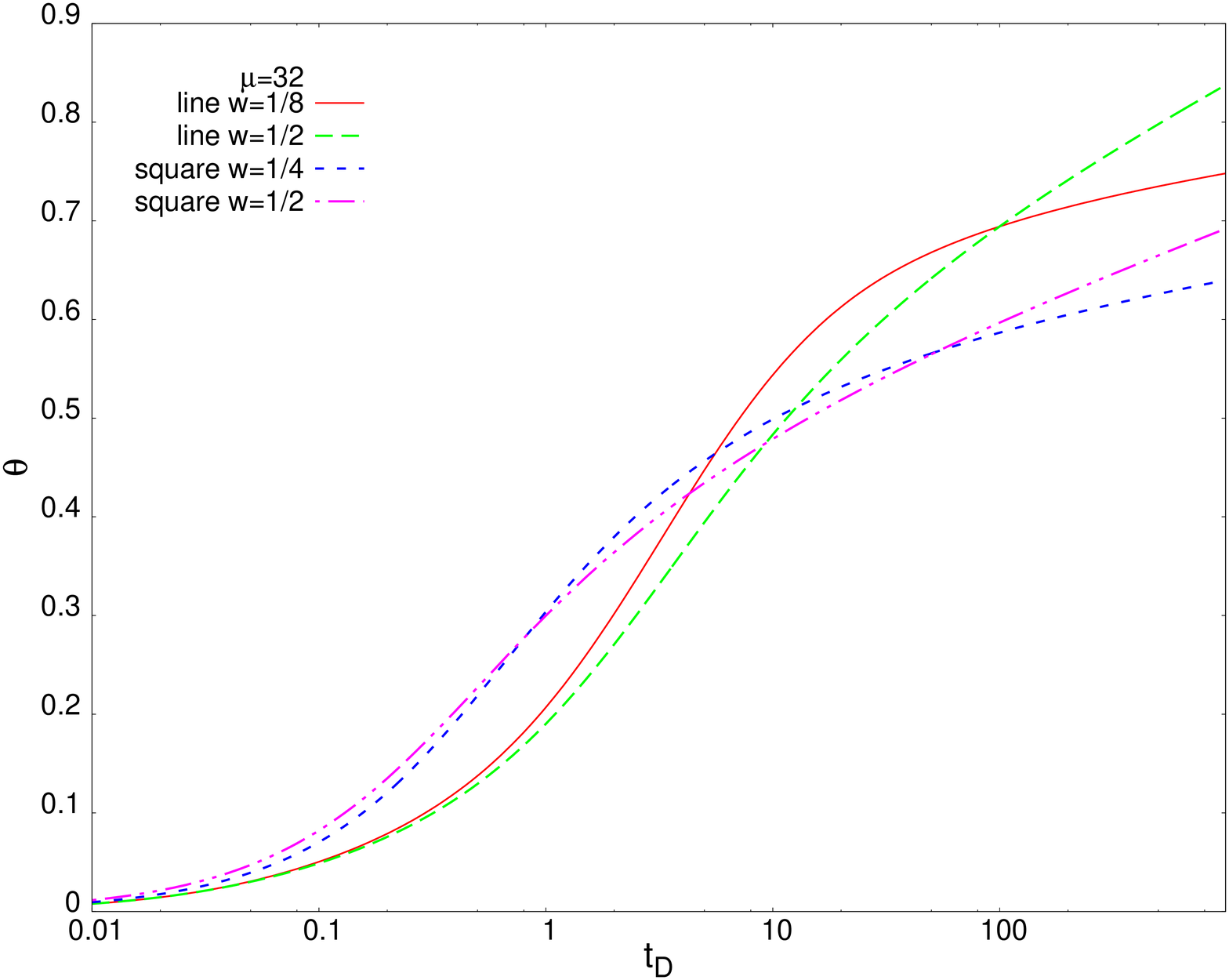}
\caption{(Color online) Comparison of the coverage evolution in linear and square particle RSA
with average incident size $\mu=32$ and small and large distribution widths.}
\label{comparison}
\end{figure}

Fig. \ref{comparison} compares the coverage evolution with linear and square particles
with the same incident size distributions $Q\left( s\right)$.
At short times ($t_D\sim 1$), $\theta$ is larger with squares.
In this regime, most of the substrate is empty, thus the adsorption of a single square of
$s^2$ particles is highly probable and instantaneously blocks a region of lateral size $s$
for future adsorption.

On the other hand, in linear particle adsorption, the same increase of the coverage in the
same region requires adsorption of $s$ aligned particles of size $s$, which occurs
after rejection of several adsorption attempts (e. g. rejection of particles aligned in the
orthogonal direction).
At long times ($t_D\sim 10$ or larger), the opposite trend is observed: linear particle
deposition gives larger $\theta$.
In this case, it is easier for several linear particles to find room for
adsorption if compared to a square of the same lateral size.

\subsection{Extrapolations of surface coverage}
\label{squarecoveragemore}

The simple exponential decay of Eq. (\ref{thetainf}) is also expected to describe the
coverage variation at long times because the square particle RSA is consistent with the
theoretical approach of Ref. \protect\cite{cornette2007}.
However, the deviations are larger than those of linear particle adsorption up to $t_D=1000$,
which indicates the presence of huge scaling corrections to the form in Eq. (\ref{thetainf}).

% new paragraph R2Q4
Despite these concerns, we estimated the relaxation times $t_R$ from linear fits
of $\log{\left( 1 -\theta\right)}\times t_D$ and show them in Table \ref{trsquare}.
We observe that $t_R$ increases with $\mu$ and decreases with $w$, which is again consistent
with a faster relaxation when the flux of monomers and other small particles increases.
However, all tabulated values of $t_R$ are larger than the maximal simulated time,
which suggests the possibility of deviations at much longer times.

\begin{table}[!h]
\centering
\label{trsquare}
\begin{tabular}{|r|c|c|c|c|}
\hline
$\mu \backslash  w$ & 1/16  & 1/8   & 1/4  & 1/2  \\ \hline
8           &       & 12500 & 5160 & 2120 \\ \hline
16          & 25200 & 13900 & 7190 & 4220 \\ \hline
32          & 25600 & 15200 & 8450 & 5870 \\ \hline
64          & 27800 & 15700 & 9200 & 6900 \\ \hline
\end{tabular}
\caption{Relaxation time $t_R$ of square particle RSA.}
\end{table}

Figs. \ref{coveragesquare}a-b show that the average size $\mu$ has a small effect on the
coverage if the ratio $w$ is constant and $t_D\leq 10$.
The general trend is $\theta$ to decrease as $\mu$ increases, for fixed $t$ and $w$.
Comparison of data for constant $\mu$ and varying $w$ (Figs. \ref{coveragesquare}c,d)
also show a small effect of $w$ up to $t_D\sim 100$.
These results contrast with the RSA of linear particles, in which a significant dependence
of the coverage on $\mu$ and $w$ was observed at $t_D\sim 1$ or longer.

% comment removed R2Q6
For fixed time and fixed $w$, the dependence of $\theta$ on $\mu$ also fits Eq. (\ref{thetan}),
with $\theta_n\left( t\right)$ interpreted as the coverage of infinitely large squares at time $t$.
Table \ref{tablesquare} shows the estimates $\theta_n$ obtained in the
extrapolations performed for various $w$ and times $t_D=10$ and $t_D=100$.
The fitting exponents $\alpha$ (Eq. \ref{thetan}) are also shown in Table \ref{tablesquare}.
The small variation of $\theta$ with $\mu$ indicates that results for the largest size $\mu =64$
represent quantitatively the RSA of squares with much larger sizes.
This is an important result to justify the generalization of our estimates of $t_S$ and
$\theta_S$ to any $w$ and $\mu$.

\begin{table}[!h]
\centering
\label{tablesquare}
\begin{tabular}{|r|c|c|c|c|c|c|}
\hline
$t_D$                        & \multicolumn{3}{c|}{10} & \multicolumn{3}{c|}{100} \\ \hline
$w$                          & 1/2   & 1/4   & 1/16   & 1/2    & 1/4   & 1/16   \\ \hline
$\theta_n$                   & 0.4634 & 0.4815 & 0.4840 & 0.5719 & 0.5823 & 0.5550 \\ \hline
$\alpha$                   & 1.01 & 1.06 & 0.99 & 1.05 & 1.04 & 0.98    \\ \hline
$\theta\left(\mu=64\right)$ & 0.4709 & 0.4878 & 0.4911 & 0.5837 & 0.5918 & 0.5636 \\ \hline
$\Delta\theta$ (\%)          & 1.6 & 1.3 & 1.4 & 2.0 & 1.6 & 1.5 \\ \hline
\end{tabular}
\caption{Estimated coverage $\theta_n$ of infinitely large squares, fitting exponents $\alpha$,
coverage $\theta\left(\mu=64\right)$ of the largest simulated square particles, and relative
difference $\Delta\theta$ between those coverages for the indicated distribution widths $w$
and times $t_D$.}
\end{table}

% paragraph changed R2Q6
Estimates of jamming coverages in lattice RSA of squares of fixed size (monodisperse)
are provided in Refs. \protect\cite{schaaf1988,nakamura,kriuch}:
for size $\mu=30$, it is $\theta_{jam}=0.574$, and for infinitely large squares, extrapolations
give $\theta_{jam}=0.564$.
This shows a small dependence on the size $\mu$, similarly to our results.
In RSA of parallel squares in the continuum, the jamming coverage is $\theta_{cont}=0.562$ \cite{brosilow}.
At $t_D=100$, our estimates of $\theta_n$ for $w=1/4$ and $w=1/2$ exceed the jamming limit of
infinite squares in the lattice and of parallel squares in the continuum.
For $w=1/16$, our estimate is slightly smaller than that value, but the difference is below $2\%$.
Thus, even with small polydispersity, a coverage near the jamming limits of lattice or continuum squares
of fixed size is reached at times of order $100\tau$.

\subsection{Adsorbed particle size distribution}
\label{squaredistribution}

\begin{figure}[!h]
\centering
\includegraphics[width=9cm]{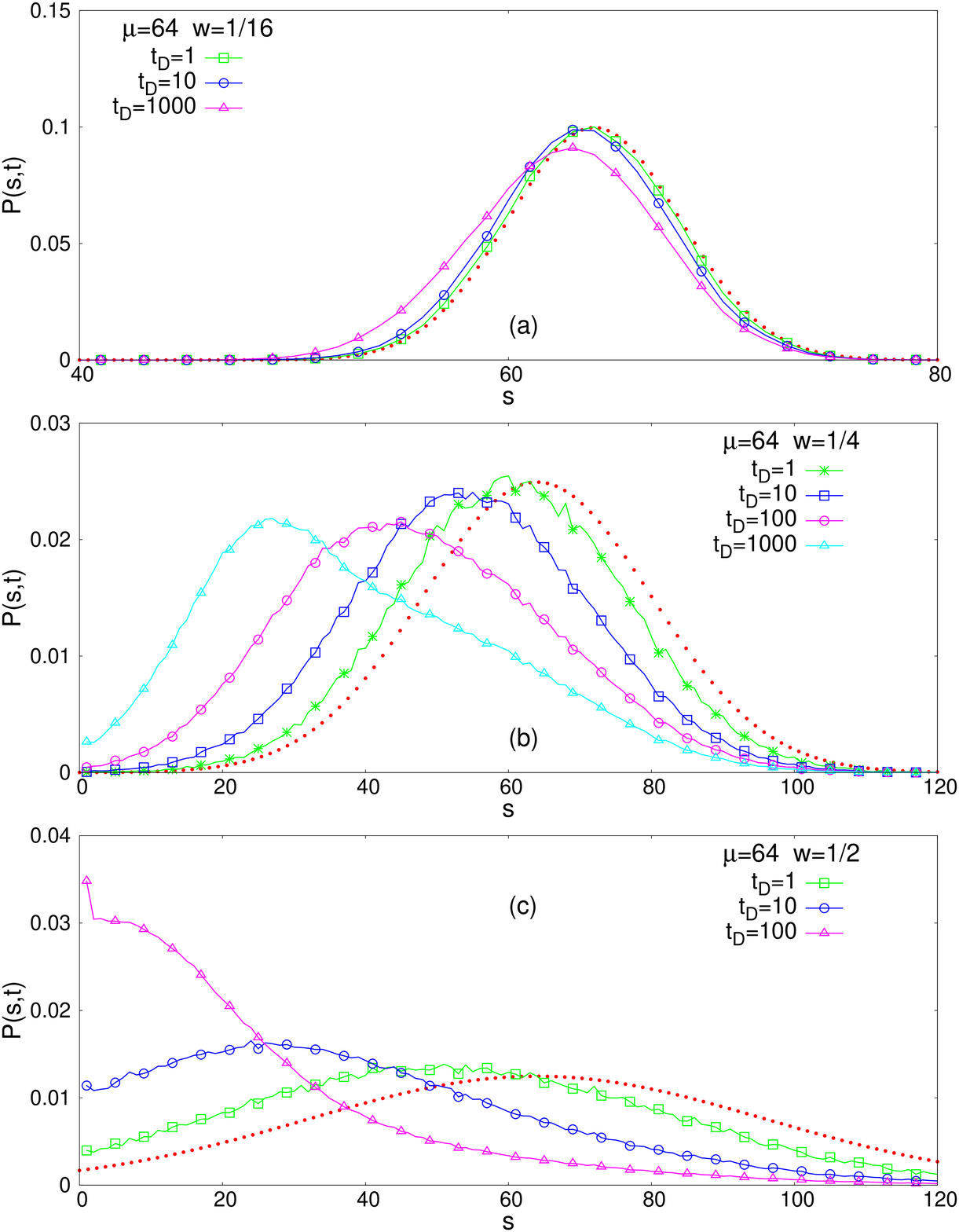}
\caption{(Color online)
Solid curves are adsorbed distributions of square particles with incident average size
$\mu=64$ and several values of width $w$ and time $t_D$. The dotted curves are the incident
distributions $Q\left( s\right)$.}
\label{distrsquare}
\end{figure}

Figs. \ref{distrsquare}a-c show the evolution of the adsorbed distribution
for $\mu =64$ and three values of $w$.
The distributions for smaller values of $\mu$ are approximately the same.

% small change R2Q7
For a small width ($w=1/16$; Fig. \ref{distrsquare}a), the difference from the incident
distribution is small up to long times.
At $t_D=1000$, only a small shift of the peak to $s<\mu$ is observed;
the coverage at this time is $\theta\approx 0.6$.
Again, the shape of the adsorbed distribution may change at much longer times due to the very
slow adsorption of small particles.
For width $w=1/4$ (Fig. \ref{distrsquare}b), a small shift of the distribution peak is
observed at $t_D=1$.

The shift is enhanced at longer times and broadening of the distribution is clear at $t_D=100$.
For the largest width ($w=1/2$; Fig. \ref{distrsquare}c), a significant shift of the distribution
peak is observed at $t_D\approx 1$, but it still has a shape similar to a Gaussian;
for $t_D\approx 10$, the monomer peak ($s=1$) is formed, and for $t_D\sim 100$,
the distribution is monotonically decreasing.

In RSA of disks with Gaussian radius distributions \cite{meakin}, a different shape of
adsorbed distribution is observed for large width-to-average ratio.
For instance, for $w=0.175$ and long times, the initial peak is shifted to $s$ slightly smaller
than the average radius $\langle R\rangle$ and a second, higher peak is developed at $s\sim R/3$.
This peak is present because the available space between the first adsorbed disks
can be filled by other disks whose radii are not much smaller than the average.
This is a particular feature of particles with rounded shapes.
On the other hand, in the RSA of squares on a lattice, the excluded volume condition has
a much more drastic effect, so that only monomers ($s=a$) and other small squares can be adsorbed
in a dense region.
This preferential adsorption of the minimum-sized particles was formerly observed in continuum
models with uniform incident distributions \protect\cite{adamczyk1997,danwanichakul}.

The adsorbed size distribution at $t_S$ is similar to the incident distribution for all values
of $\mu$ and $w$ up to $w=1/4$ (recall that $t_S<1$ for squares; Eq. \ref{tSsquare}).
Broadening and formation of the peak at the monomer size are observed only for times
much longer than $t_S$ or for larger $w$ (e. g. $w=1/2$; Fig. \ref{distrsquare}c).

\subsection{Pair correlation function}
\label{squarecorrelation}

\begin{figure}[!h]
\centering
\includegraphics[width=9cm]{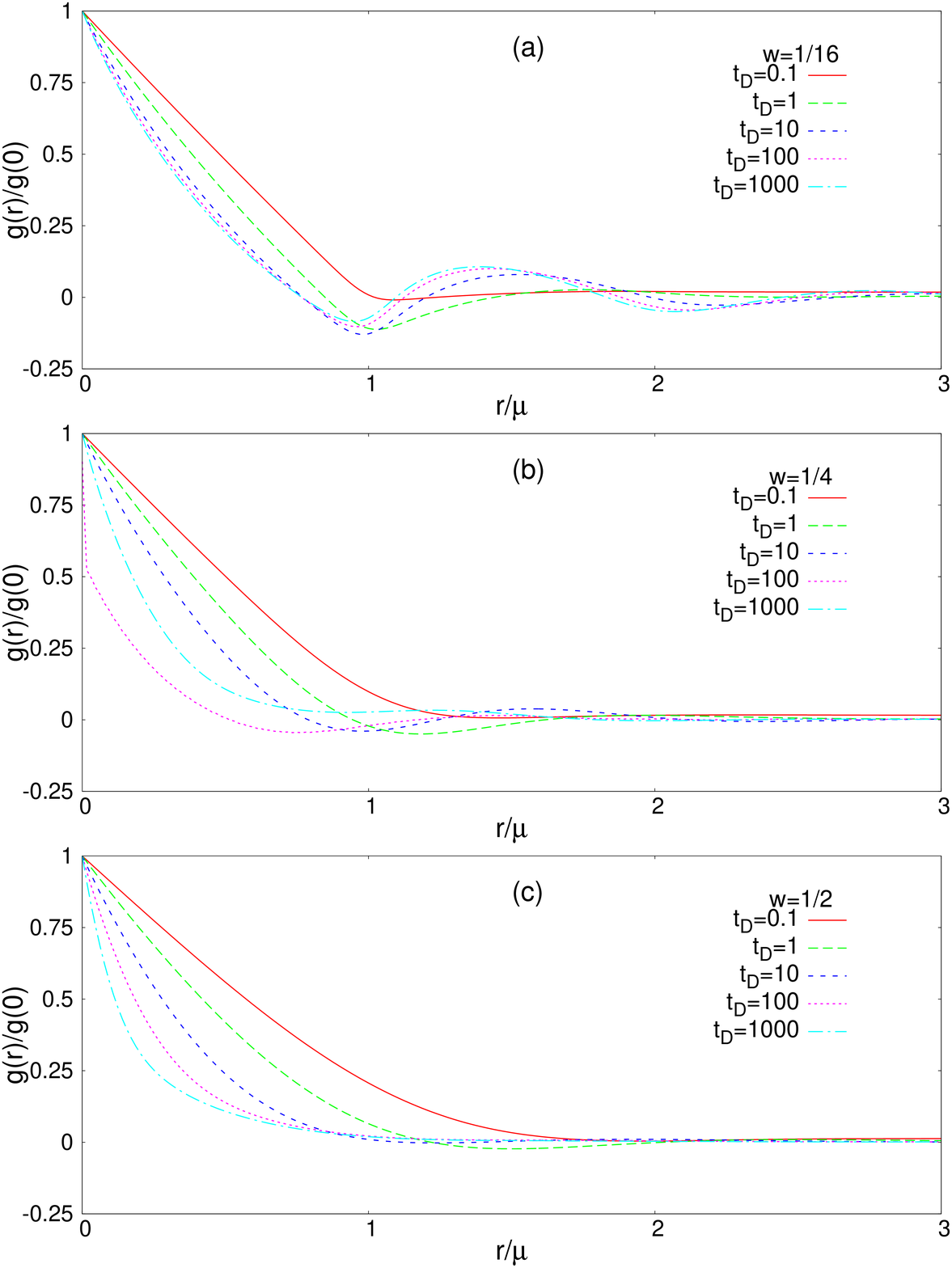}
\caption{(Color online) Scaled pair correlation as function of the scaled distance for square
particles with average incident size $\mu=64$ and several values of width $w$ and time $t_D$.}
\label{gsquare}
\end{figure}

Figs. \ref{gsquare}a-c show $g\left( r,t\right)/g\left( 0,t\right)$ as function of the reduced
distance $r/\mu$ for $\mu=64$, three ratios $w$, and several times.
The results for smaller values of $\mu$ are approximately the same.

For $w=1/16$ (Fig. \ref{gsquare}a), a mininum of $g$ begins to form at
$t_D\approx 1$ and is deeper at $t_D=10$, which is one order of magnitude longer than $t_S$.
As $t$ increases, the depth of this minimum is reduced; this means that ordering of neighboring
filled and empty regions is reduced.
On the other hand, the maximum of $g$ increases in time and is located at $r\approx1.5\mu$;
this occurs because squares of size $\approx \mu$ are separated
by distances ranging from $0$ to $\mu$, whose average is $\mu/2$, similarly to the case of
linear particles.

For $w=1/4$ (Fig. \ref{gsquare}b), a shallow minimum of $g$ is observed at $t_D=1$.
The maximum is also reduced due to the adsorption of particles of smaller sizes, particularly
at long times.
For $w=1/2$, the oscillatory structure of $g$ disappears.

Since $t_S<1$, a minimum of $g$ may be observed only for very small $w$  at $t_S$ (e. g. $w=1/16$;
Fig. \ref{gsquare}a).
The clear oscillatory structure of $g$ may observed for $w\leq 1/8$, but at times of
order $10 t_S$ and longer.

\section{Discussion and Conclusion}
\label{conclusion}

We studied RSA of particles on square lattices with incident sizes following discretized
Gaussian distributions and with conditions of no diffusion and no desorption..
The RSA of linear particles and square particles were separately considered.

We observed that plots of coverage as function of the logarithm of time are suitable to
determine the order of magnitude of the monolayer deposition time $\tau$.
Those plots have maximal slopes $S$ at real time $t_M\sim \tau$ (dimensionless time $t_S\sim 1$),
similarly to the exactly solvable case of RSA of monomers.
A detailed analysis of the peaks of $S$, including effects of particle size and polydispersity,
shows that $1.5\tau< t_M < 5\tau$ for linear particles and $0.3\tau < t_M < \tau$ for square
particles.
These particle shapes may be viewed as limiting cases of the cross section of a large
variety of nanostructures.
From the estimates of $\tau$, the orders of magnitude of molecular fluxes and/or sticking
coefficients may also be estimated.

Extrapolations of coverage data show that results obtained for average linear particle size
$\mu=64$ are qualitatively similar to those expected in much larger sizes, with possible
discrepancies up to $20\%$.
For square particles, results for the largest studied particle size $\mu=64$ are quantitatively
reasonable for any larger size.
This suggests that square particle results may used as a first approximation in adsorption studies
of particles with other compact shapes and aspect ratios close to $1$.

The adsorbed particle size distributions slowly change in time for small relative
widths $w$ of the incident distribution ($w\leq 1/8$) with both particle shapes.
At long times ($t\sim 100\tau$ or longer) and $w=1/4$, a significant shift of the distribution
peak and broadening are observed.
For $w=1/2$, a monotonically decreasing distribution is observed at $t\sim 100\tau$ because
most adsorbed particles are small.

Comparison of incident and adsorbed distributions at $t_S$ (maximum of $S$) shows that
they are similar for $w\leq 1/8$.
For $w=1/4$, the adsorbed distribution is slightly shifted to the left, and for $w=1/2$
there are significant differences in those distributions even at short times..
The reliability of the model for a given application may be tested by checking
the consistency with these features.

The pair correlation function $g\left( r,t\right)$ is also able to distinguish small and
large widths $w$ in the RSA of linear and square particles.
For small incident widths ($w\leq 1/8$), the minimum of $g$ is beginning to be formed
at $t_S$.
A clear oscillatory structure, with a minimum at $r\approx \mu$ and maximum at
$r\approx 1.5\mu$, is observed at times typically of order $10 t_S$.
For linear particles, this structure is enhanced at much longer times due to alignment effects,
but the opposite trend is observed with square particles.

% small changes R1Q1
Previous works have analyzed effects of polydispersity in RSA models.
Some of them considered Gaussian distributions of incident particle size
\cite{meakin,adamczyk1997,marques,hanarp} in off-lattice RSA.
However, the present work is not a simple lattice extension of those RSA models of Gaussian
mixtures.
It also advances by showing how the properties of the coverage evolution can be combined with
the study of other quantities (e. g. size distributions and correlation functions) to estimate
characteristic times and sizes and, possibly, physico-chemical parameters such as molecular
flux and sticking probabilities.
For this purpose, our analysis focused on the model features at times of order $\tau$ instead
of long time features.

We believe that our approach may be useful for future experimental work because it suggests a
procedure to identify the characteristic time $\tau$ and shows how the adsorbate features at
times of this order are related to polydispersity.
Our results may also be useful in future RSA studies, for instance to models including particle
interactions, particle mobility, and other particle shapes and substrate structures \cite{cook,herman}.
The detailed study of features at times of order $\tau$ may also be interesting in off-lattice
(continuum) RSA, for instance considering the variety of particle shapes studied in recent
works for modeling compact and branched structures \cite{ciesla2014,ciesla2015}.

\acknowledgments

The authors acknowledge support from CNPq and FAPERJ (Brazilian agencies).

%~~~~~~~~~~~~~~~~~~~~~~~~~~~~~~~~~~~~~~~~~~~~~~~~~~~~~~~~~~~~~~~~~~~~~~~~~~~
%~~~~~~~~~~~~~~~~~~~  REFERENCES  ~~~~~~~~~~~~~~~~~~~~~~~~~~~~~~~~~~~~~~~~~~
%~~~~~~~~~~~~~~~~~~~~~~~~~~~~~~~~~~~~~~~~~~~~~~~~~~~~~~~~~~~~~~~~~~~~~~~~~~~

\end{document}